\documentclass[prl,twocolumn,amsmath,amssymb,floatfix,superscriptaddress,notitlepage,nofootinbib]{revtex4-1}

\usepackage{amssymb,amsmath,framed}
\usepackage[dvipdfmx]{graphicx,color}
\usepackage{xcolor}

\begin{document}

\title{Invertible field transformations with derivatives: necessary and
sufficient conditions}

\author{Eugeny Babichev}
\affiliation{Laboratoire de Physique Th\'eorique (UMR 8627), CNRS, Univ. Paris-Sud, Universit\'e Paris-Saclay, 91405 Orsay, France}

\author{Keisuke Izumi}
\affiliation{Kobayashi-Maskawa Institute, Nagoya University, Nagoya 464-8602, Japan}
\affiliation{Department of Mathematics, Nagoya University, Nagoya 464-8602, Japan}

\author{Norihiro Tanahashi}
\affiliation{Institute of Mathematics for Industry, Kyushu University, Fukuoka 819-0395, Japan}

\author{Masahide Yamaguchi}
\affiliation{Department of Physics, Tokyo Institute of Technology,
2-12-1 Ookayama, Meguro-ku, Tokyo 152-8551, Japan}

\date{\today}

\begin{abstract}
We formulate explicitly the necessary and sufficient conditions for the
local invertibility of a field transformation involving derivative
terms.  Our approach is to apply the method of characteristics of
differential equations, by treating such a transformation as
differential equations 
that give new variables in terms of  original ones. The obtained results
generalise the well-known and widely used inverse function theorem.
Taking into account that field transformations are ubiquitous in modern
physics and mathematics, our criteria for invertibility will find
many useful applications.
\end{abstract}

\maketitle%

\section{Introduction}

Transformations of fields (or variables) are very
common in physics and mathematics. 
From one hand it provides better ways to understand various physical phenomena, 
and on the other hand --- to advance calculations, in particular, to solve more easily differential equations.
A change of gauge and/or a global redefinition of fields is a popular
example of a field transformation. 
Some physical phenomena can be more
easily understood in a particular gauge or frame. 
The decomposition of a complex field (more generally, multiple fields transforming covariantly under a gauge or a global transformation) into radial and phase
directions is also useful to identify the Nambu-Goldstone mode
\cite{Nambu:1960tm,Goldstone:1961eq} after a symmetry breaking, and 
such a decomposition can be regarded as a field transformation. Another
example is a Bogoliubov transformation
\cite{Bogoljubov:1958,Valatin:1958}, which enables us to investigate 
effects of the change of vacua, in particular, the Hawking radiation.
A Fourier (including a Fourier series expansion) 
and a Laplace transformation are also regarded as field
transformations and are often used in physics and mathematics. A Galilean
transformation, a Lorentz transformation, and a general coordinate
transformation lead to the transformation of a tensor field. 
In gravity, a conformal transformation is frequently used to change frames from the Jordan
one to the Einstein one and vice versa, and also to identify the degrees of
freedom of a gravitational field
\cite{York:1971hw,York:1972sj}. Recently, a more general transformation
called a disformal transformation \cite{Bekenstein:1992pj} 
has been extensively used to change the frame or to relate
different gravity theories~\cite{Zumalacarregui:2013pma}. Thus, field transformations are
ubiquitous in physics and mathematics.

Among them, ``invertible'' field transformations are particularly
important because they do not change the number of dynamical degrees of
freedom and yield equivalent equations. 
For a field transformation without derivative terms
$\tilde{\phi}_i(x^{\mu}) = \tilde{\phi}_i[\phi_j(x^{\mu})]$,
the condition for the (local) invertibility is
given by the celebrated inverse function theorem, that is, $\det(\partial
\tilde{\phi}_i/\partial \phi_j) \ne 0$. 
However,  
more general transformations 
such as a disformal transformation and a gauge transformation
may involve derivatives of fields.
For example, a disformal transformation
\cite{Bekenstein:1992pj} is given by $\tilde{g}_{\mu\nu} =
A(\phi,X)g_{\mu\nu} + B(\phi,X)\partial_{\mu}\phi\partial_{\nu}\phi$
with $X = g^{\sigma\tau}\partial_{\sigma}\phi
\partial_{\tau}\phi$.
However, to the best of our knowledge, no  
criterion for the (local)
invertibility of a field transformation involving derivatives
has been given yet. 
The purpose of this {\it Letter} is to give an explicit formulation of such a transformation\footnote{Equivalence of the equation of motion for
a given invertible transformation with derivative terms is shown in Ref.~\cite{Takahashi:2017zgr}, 
whereas criteria for invertibility of the transformation are not discussed therein.}.

Our strategy to find the invertibility condition for a transformation with derivatives
is to use the method of characteristics for differential equations~\cite{CH:1962,Ong:2013qja,Izumi:2013poa,Izumi:2014loa,Reall:2014pwa}. 
We treat field transformation with derivative terms as 
differential equations 
that give new variables in terms of old variables.
We then investigate properties of these differential equations, and find
the necessary conditions for invertibility.
After that, we show that they are also sufficient conditions. 
Throughout this work we primarily focus on the simplest non-trivial transformations involving two fields and their first-order derivatives. 
However, our method applies also to transformations with 
any (finite) number of fields and their derivatives of any (finite) order, 
which will be discussed in future publications. In the following, we will first
explain our approach in some detail, and then derive the necessary and
sufficient conditions for the (local)   invertibility. 
We also give non-trivial examples of invertible transformations
to demonstrate how our criteria for invertibility apply and how powerful they are. 
Finally, we will summarise and discuss our results.

\section{Necessary conditions}

We consider a transformation from a theory of $\phi_i(x^\mu)$ to that of
$\psi_a(x^\mu)$ and its derivatives%
\footnote{Let us comment here on a
possible confusion related to a gauge transformation. For example,
performing the gauge transformation of a vector field $\tilde{A}_\mu =
A_\mu + \partial_\mu\alpha$, where $\alpha$ is a gauge function, one
should think of $\alpha$ as a part of both sets of fields,
i.e.\ $\{A_\mu,\alpha\}$ and $\{\tilde{A}_\mu,\alpha\}$.  In this case
the number of fields does not change and this transformation falls into
the class we consider in this {\it Letter}. }:
\begin{equation}
\phi_i=\phi_i\left( \psi_a, \partial_\mu, x^\mu \right).
\label{phipsi1}
\end{equation}
Note that $\phi_i$ and $\psi_a$ are functions, that is, they can be scalars and/or components of vectors and tensors.

Here we will sketch the derivation of the invertibility conditions for
the transformation~(\ref{phipsi1}), assuming the numbers of the fields $\phi_i$ and $\psi_a$ are both $n$. 
For the transformation to be invertible, the number of (independent)
fields $\phi_i(x^\mu)$ and $\psi_a(x^\mu)$ should be the same,
because, otherwise, the degrees of freedom before and after the
transformation would not coincide.
Another (necessary) condition we require
is {\it absence of the characteristics}\footnote{We will understand characteristics in a
broader sense, including also complex-valued solutions of the characteristic equation.
See the discussion below.} 
when regarding Eq.~(\ref{phipsi1}) as evolution equations for $\psi_a$ in
terms of $\phi_i$. (See e.g.\ Ref.~\cite{CH:1962} for the basics of
characteristics.)  By definition, a characteristic surface (or simply
put, a characteristic) is defined as a surface at which the evolution
equation becomes singular, in the sense that the coefficient of the
highest derivative term vanishes.
At such a surface, the solution $\psi_a(\phi_j)$ of the differential equations~(\ref{phipsi1}) becomes non-unique,
implying non-invertibility.  Notice that absence of
characteristics of Eq.~(\ref{phipsi1}) is a necessary but not a sufficient
condition.   

 To analyse characteristics, we consider a $(D-1)$-dimensional hypersurface in $D$-dimensional spacetime, and let $\xi^\mu$ be a vector in a non-tangential direction for the surface.
Then the structure of characteristics of Eq.~(\ref{phipsi1}) is governed
by a $n\times n$ matrix $M$ defined as
\begin{equation}
M_{ia}(\xi^\mu)
:=
\frac{
\partial \phi_i
}{\partial \left(
\partial_{\mu_1}\cdots \partial_{\mu_{N(i,a)}} \psi_a
\right)}
\xi^\mu \cdots \xi^{\mu_{N(i,a)}}
,
\label{M}
\end{equation}
where 
$N(i,a)$ is the order of the highest derivative term of $\psi_a$
in the $i$th component of Eq.~(\ref{phipsi1}). 
This matrix is called the characteristic matrix.
The discussion of characteristics makes
sense only for quasi-linear equations, while Eq.~(\ref{phipsi1}) is
nonlinear in $\partial^{N(i,a)} \psi_a$ in general.  To convert
(\ref{phipsi1}) into a quasi-linear differential equation of
$\psi_a$, we differentiate it with respect to $x^\mu$ sufficient number of times.  
As a result of the differentiation additional characteristics are
introduced, while they can be easily separated from the inherent
ones.

For simplicity, we assume that orders of derivatives are $N$
(ignoring the derivatives we applied above)
for all the components of (\ref{phipsi1}), while
generalisation is straightforward.
We also define the number of derivatives of Eq.~(\ref{phipsi1}) as $Nn$.
To avoid  characteristics associated with the
derivatives in (\ref{phipsi1}), the determinant of the characteristic
matrix $M$ must be identically zero:
\begin{equation}
\det  M(\xi^\mu) \equiv 0, 
\label{M}
\end{equation} 
because, otherwise, this equation would have particular solutions
$\xi^{\mu}$ yielding  
characteristics. 
This condition
implies that the degree of degeneracy of $M$ is $m>0$.
Then, taking linear combinations appropriately, (the differentiated) Eq.~(\ref{phipsi1}) can be reduced to
a system of $(n-m)$ equations with $N$th (inherent) derivatives and $m$ equations with $(N-1)$th (inherent) derivatives.
Let us define the characteristic matrix for the new system of equations as $M'$.
If $Nn-m \neq 0$, the (inherent) characteristics are present and the transformation is not invertible.
Therefore, for the transformation to be invertible, extra degeneracy should be present,
that is, $\det\bigl( M'(\xi^\mu) \bigr)$ must be identically
zero also. 
We should repeat this procedure until the total degree of the degeneracy becomes equal to $Nn$. 
We define the characteristic matrix at this point as $\hat M$. 
Now, the order of $\det \hat M$ with respect to $\xi^\mu$ is the same as the number of the derivatives we applied,
and all inherent derivatives of Eq.~(\ref{phipsi1}) are removed. 
If $\hat M$ is degenerated, Eq.~(\ref{phipsi1}) implicitly gives a relation among $\psi_i$'s, and hence it is not invertible.
Therefore, for any $\xi^\mu$ which does not correspond to the direction of the derivatives we applied to Eq.(\ref{phipsi1}), $\det \hat M (\xi^\mu) \neq 0$ is required --- in fact, it corresponds to the condition for the inverse function theorem.
This is the procedure to obtain the necessary conditions for the invertibility of a transformation (\ref{phipsi1}).

To see physical implications of an invertible transformation, let us
apply the transformation
(\ref{phipsi1}) to the equations of motion for $\phi_i$ given by
\begin{align}
F_j \left(\phi_i, \partial_\mu, x^\mu \right) =0.
\label{Fj}
\end{align} 
Applying (\ref{phipsi1}) yields equations of motion in terms of $\psi_a$.
Rather than solving them in terms of $\psi_a$, 
we instead can regard 
Eqs.~(\ref{phipsi1}) and (\ref{Fj}) as a system of differential equations for $\psi_a$ with auxiliary fields $\phi_i$.
The structure of the characteristics matrix for this system of equations is
\begin{align}
\bordermatrix{     & \phi_i & \psi_a  \cr
              \mbox{Eq.~(\ref{Fj})}& M_{11} &  0  \cr
               \mbox{Eq.~(\ref{phipsi1})} & M_{21} & \hat M 
            } \ . \label{Mat}
\end{align} 
Here, $\hat M$ is the characteristic matrix for (\ref{phipsi1}) with the degenerate part   removed by the procedure explained above.
The determinant of (\ref{Mat}) becomes
$\det  M_{11} \times \det \hat M$.
As we noted above, $\det \hat M \neq 0$ must be satisfied for an invertible transformation.
Hence, 
the characteristic equation ($\det M_{11}\times\det\hat M=0$) reduces to
$\det M_{11}=0$.
The characteristics in this theory are given as surfaces satisfying this equation, and 
all of them coincide with those in the theory of $\phi_i$ with Eq.~(\ref{Fj}). 
Hence we have confirmed that an invertible transformation keeps characteristics in the original theory invariant.

Let us here comment on the case where there are no real characteristics while the order of Eq.~(\ref{M}) in $\xi^\mu$ is non-zero.
This occurs when Eq.~(\ref{phipsi1}) is a non-hyperbolic partial differential equation and Eq.~(\ref{M}) has no real-valued solutions.
Even in this case, as long as $\det M$ is non-zero Eq.~(\ref{phipsi1}) will have solutions specified by parameters corresponding to integration constants, at least locally.
This implies that 
the fields $\psi_a$ cannot be expressed uniquely in terms of $\phi_i$ and the transformation is not invertible.
Thus, for the transformation to be invertible, it is necessary that the coefficients of the highest-order derivatives degenerate completely, that is, $\det M$ is identically zero. 
This also applies to $M'$ and other matrices in the construction sketched above, when we required that the determinant of a matrix was identically zero.

\section{Simplest case}
Let us apply the strategy above to the simplest (but non-trivial) case for a transformation from two fields $\phi_i$ $(i=1,2)$ to other two fields $\psi_a$ $(a=1,2)$ including the first-order derivatives.
The transformation (or say the field redefinition) can be written as 
\begin{align}
\phi_i=\phi_i\left( \psi_a, \partial_\alpha \psi_a  \right) \qquad (i=1,2,\quad a=1,2). \label{phipsi}
\end{align}
We define two matrices
\begin{align}
A^\mu_{ia}:= \frac{\partial \phi_i}{\partial (\partial_\mu \psi_a)}, \qquad 
B_{ia}:= \frac{\partial \phi_i}{\partial \psi_a}.
\end{align}
If $A^\mu_{ia}=0$, the transformation does not involve derivatives.
In this case we can apply the standard inverse function theorem and the invertibility follows if $\det(B_{ia})\neq 0$.

Let us focus on the non-trivial case $A^\mu_{ia}\neq 0$ henceforth.
The number of derivatives in this system is two, 
and thus, 
degeneracy of two degrees is required.
If 
the transformation~(\ref{phipsi}) is
 nonlinear for the first-order derivatives, 
we apply an extra derivative to them. 
Then, the obtained (second order) equations are quasi-linear. 
Their characteristics consist of those of the original equation (if present) and the extra characteristics originated from derivatives we applied to convert Eq.~(\ref{phipsi}) in quasi-linear form. 
We can easily discriminate them from inherent derivatives.

The degeneracy required for the invertibility reads,
\begin{eqnarray}
0\equiv\det \left( A^\alpha_{ia}\xi_\alpha \right)
\end{eqnarray} 
for arbitrary $\xi^\mu$, or equivalently
\begin{eqnarray}
0\equiv\epsilon_{i_1i_2} \epsilon_{a_1a_2} A^{(\alpha_1}_{ i_1 a_1} A^{\alpha_2)}_{ i_2a_2} .\label{L}
\end{eqnarray}
The number of degeneracy from the above condition should be one, 
otherwise $A^\mu_{ia}= 0$. 
Thanks to this degeneracy, we can find one combination of the (differentiated) transformation~(\ref{phipsi}) that is one order lower in derivatives (details will be given in the forthcoming paper).
For the sake of the invertibility, this equation must be degenerate by one order.
Such a requirement results in
\begin{eqnarray}
{\cal A}^{(\alpha}_{2,ab} {\bar A}^{\beta)}_{bj}\equiv0, \label{SL}
\end{eqnarray} 
where
\begin{equation}
{\bar A}^{\alpha}_{ai}:=  \epsilon_{i j}\epsilon_{ab} A^{\alpha}_{jb}, 
~~~
{\cal A}^{\alpha}_{2,ab}:=  {\bar A}^{\alpha}_{aj} B_{jb} +  {\bar A}^{\beta}_{aj}\left(\partial_\beta  A^{\alpha}_{jb}\right) .
\end{equation}
Now we have degeneracies for two orders, 
which are enough 
to establish the invertibility in our case.
The resultant equations, on the other hand, should not be degenerate.
This requirement results in
{\begin{align}
&
\left\{
\left[
{\cal A}_{2,ab}^\alpha { A}_{ib}^{\mu_1} A_{ic}^{\mu_2}
-{\cal A}_{2,ab}^{\mu_1} 
\left(
{A}_{ib}^{\mu_2} A_{ic}^\alpha + {\bar A}_{bi}^\alpha {\bar A}_{ci}^{\mu_2} 
\right)
\right]
\left( \partial_\alpha {\bar A}_{cj}^{\mu_3} \right) 
\right. \nonumber \\
&\left. 
+
\left(
A^{\mu_1}_{kb}A^{\mu_2}_{kb} 
\bar A_{ai}^\beta \partial_\beta B_{ic}
- {\cal A}^{\mu_1}_{2,ab} A^{\mu_2}_{ib}   B_{ic}
\right)
\bar A^{\mu_3}_{cj}
\right\} 
\xi_{\mu_1}\xi_{\mu_2}\xi_{\mu_3}
 \neq 0
\label{SSL}
\end{align}
for arbitrary $\xi^\mu$.

The conditions (\ref{L}), (\ref{SL}) and (\ref{SSL}) can be simplified further.
The condition~(\ref{L})
fixes the form of $A_{ij}^\mu$ as
\begin{eqnarray}
A_{ia}^\mu = a^\mu V_i U_a, \label{aVU}
\end{eqnarray}
where $V_i$ and $U_a$ can be normalized as
\begin{eqnarray}
V_iV_i=1, \qquad U_aU_a=1. \label{norm}
\end{eqnarray}
Then, the conditions (\ref{SL}) and (\ref{SSL}) can be written as
\begin{gather}
 n_i B_{ia} m_a \equiv0, \label{nBm}\\
 n_i B_{ia}U_a\neq 0, \quad   \left( V_i B_{ia} -  a^\beta \partial_\beta U_a \right) m_a \neq 0, \label{nBU}
\end{gather}
where $n_i:=\epsilon_{ij} V_j$, $m_a := \epsilon_{ab} U_b$.
In the above, $a^\mu$, $V_i$, $U_a$, $B_{ia}$, $n_i$ and $m_a$ are
functions of $\psi_b$, $\partial_\alpha \psi_b$ and $x^\mu$.
To summarise, the necessary conditions for the invertibility are found to be 
(\ref{aVU}), (\ref{nBm}) and (\ref{nBU}).
Equations~(\ref{aVU}) and (\ref{nBm}) are necessary to realize the degeneracies, and (\ref{nBU}) corresponds to $\det \hat M\neq 0$ explained in the previous section.

\section{Sufficient conditions}

In the previous section, we derived the necessary conditions (\ref{aVU}), (\ref{nBm}), (\ref{nBU}) for the transformation~(\ref{phipsi}) to be invertible.
We can prove that these conditions become also the sufficient conditions for the invertibility.
However, since the proof is lengthy, 
in this {\it Letter} we show the proof within the perturbative regime.
We will show the complete proof in the forthcoming paper.

We start from the perturbative expansion of the transformation (\ref{phipsi});
\begin{eqnarray}
\delta \phi_i = A_{ia}^\mu \partial_ \mu \delta \psi_a + B_{ia} \delta \psi_a.
\end{eqnarray}
Here, $A_{ia}^\mu$ and $B_{ia}$ are functions of $\psi_b(x^\mu)$,
$\partial_\alpha \psi_b(x^\mu)$ and $x^\mu$.
In the perturbative analysis of this section, we regard them as fixed functions and assume only $\delta \psi_a$ are dynamical variables. 
We will show the uniqueness of $\delta \psi_a$, that is,
if $\delta \psi_a^{(1)}$ and $\delta \psi_a^{(2)}$ are
the solution of the above equations for the same $\delta \phi_i$,
$\delta \psi_a^{(1)}$ is equal to $\delta \psi_a^{(2)}$. 
For this purpose
it is sufficient to show that if $\Psi_a := \delta \psi_a^{(1)}- \delta
\psi_a^{(2)} $ satisfies
\begin{eqnarray}
A_{ia}^\mu \partial_ \mu \Psi_a + B_{ia} \Psi_a=0, \label{Psi0}
\end{eqnarray}
then $\Psi_a $ is zero.

Using the condition (\ref{aVU}), Eq.~(\ref{Psi0}) is written as 
\begin{eqnarray}
a^\mu V_i U_a \partial_ \mu \Psi_a + B_{ia} \Psi_a=0. \label{Psi1}
\end{eqnarray}
Contracting $n_i$ with this equation, we have
\begin{eqnarray}
\left( n_i B_{ia} U_a \right) \left(U_b \Psi_b\right)=0,
\end{eqnarray}
where we used the condition (\ref{nBm}).  Because of the condition
(\ref{nBU}), this equation results in
\begin{eqnarray}
U_b \Psi_b=0. \label{UPsi}
\end{eqnarray}

On the other hand, applying $V_i$ to  Eq.~(\ref{Psi1}) we have 
\begin{align}
\left(V_i B_{ia}- a^\mu \partial_\mu U_a\right)m_a \left( m_b \Psi_b \right)=0,
\end{align}
where we used the condition (\ref{UPsi}).  
Combined with
(\ref{nBU}), this equation 
implies
\begin{eqnarray}
m_b \Psi_b=0. \label{mPsi}
\end{eqnarray}
As a result, we have 
\begin{eqnarray}
\Psi_a= U_a \left( U_b \Psi_b\right)+m_a \left( m_b \Psi_b\right) = 0. 
\end{eqnarray}
This means that, 
in the perturbative regime,
(\ref{aVU}),
(\ref{nBm}) and (\ref{nBU}) are the sufficient conditions for
the invertibility.

\section{Example of invertible transformation}
It is not difficult to check that our conditions are satisfied for the
simple case
$\phi_1 = \psi_1+ f(\partial_\mu\psi_2)$, $\phi_2=\psi_2$, a direct
analogue of the disformal transformation in gravity.
This case is trivially invertible, and the conditions
(\ref{aVU})--(\ref{nBU}) are indeed satisfied by nothing that $a^\mu =
\partial f/\partial \bigl(\partial_\mu\psi_2\bigr)$, $V_i = (1,0)$ and $U_a=(0,1)$.

Below we give a non-trivially invertible example, that is, whose
invertibility cannot be established without the use of the conditions
formulated in this {\it Letter}.
Let us consider 
a transformation that is linear in $\partial \psi_a$ whose coefficients are functions of $\psi_a$, that is,
\begin{align}
\phi_i = a^{\mu}(\psi_b) V_i(\psi_b) U_a(\psi_b) \partial_\mu \psi_a + f_i(\psi_b). 
\label{phipsi_linear}
\end{align}
Without loss of generality, we can set $U_a = (1,0)$%
\footnote{In Eq.~(\ref{phipsi_linear}), $U_a$ can be set to $(1,0)$ in general by a field re-definition $\psi_a = \psi_a(\psi^\text{new}_b)$ and a rescaling of $a^\mu$.  }
}, 
hence
\begin{align}
\phi_i &= a^{\mu}(\psi_a) V_i(\psi_a) \partial_\mu \psi_1 + f_i(\psi_a), 
\label{ex1}
\end{align}
This transformation 
automatically satisfies the condition~(\ref{aVU}).
The other conditions (\ref{nBm}), (\ref{nBU}) reduce to
\begin{align}
n_i \frac{\partial V_i}{\partial \psi_2} a^\mu \partial_\mu \psi_1 + n_i \frac{\partial f_i}{\partial \psi_2} \equiv 0,
\label{econd1}\\
n_i \frac{\partial V_i}{\partial \psi_1}  a^\mu \partial_\mu \psi_1 + n_i \frac{\partial f_i}{\partial \psi_1}
\neq 0,
\label{econd2}\\
\frac{\partial a^\mu}{\partial \psi_2} \partial_\mu \psi_1 + V_i \frac{\partial f_i}{\partial \psi_2}
\neq 0.
\label{econd3}
\end{align}
Since Eq.~(\ref{econd1}) must hold identically for any $\psi_a$ and $\partial_\mu\psi_a$, it follows that
\begin{equation}
n_i \frac{\partial V_i}{\partial \psi_2} \equiv 0,
\qquad
n_i \frac{\partial f_i}{\partial \psi_2} \equiv 0.
\end{equation}
The first equation implies  $V_i = V_i(\psi_1)$, and the second one guarantees that a function
$g(\psi_1)$ exists and satisfies
\begin{equation}
g(\psi_1) 
:=
n_i(\psi_1) f_i(\psi_a) 
= 
n_i \phi_i
,
\label{gdef}
\end{equation}
where the second equality follows from
(\ref{ex1}).

Using the above results, Eq.~(\ref{econd2}) becomes
\begin{equation}
\frac{\partial}{\partial\psi_1}\left[
g(\psi_1) - n_i(\psi_1) \phi_i 
\right]
\neq 0,
\label{econd2mod}
\end{equation}
where the partial derivative $\partial/ \partial \psi_1$ is taken regarding $\psi_i$ and $\phi_a$ as independent variables. 
Equation~(\ref{econd2mod}) implies the implicit function theorem can be applied to Eq.~(\ref{gdef}) to express $\psi_1$ in terms of $\phi_i$;
\begin{equation}
\psi_1 = \psi_1(\phi_i).
\label{psi1phia}
\end{equation}
Also from Eq.~(\ref{econd3}) we have
\begin{equation}
\frac{\partial}{\partial \psi_2}\left[
a^\mu(\psi_a) \partial_\mu \psi_1 + V_i(\psi_1) f_i(\psi_a)
\right]
\neq 0. \label{if2}
\end{equation}
Transformation equation (\ref{ex1}) gives
\begin{align}
V_i (\psi_1) \phi_i &= a^{\mu}(\psi_a)  \partial_\mu \psi_1 + V_i (\psi_1)f_i(\psi_a), 
\label{exV}
\end{align}
Since $\psi_1$ and $\partial_\mu \psi_1$ can be regarded as a function of $\phi_i$ and $\partial_\mu \phi_i$ thanks to Eq.~(\ref{psi1phia}), (\ref{if2}) implies that the implicit function theorem can be applied to Eq.~(\ref{exV}) to find
\begin{equation}
\psi_2 
= \psi_2\bigl(\phi_i, \psi_1(\phi_i), \partial_\mu \psi_1(\phi_i)\bigr).
\label{psi2phia}
\end{equation}
Equations (\ref{psi1phia}) and (\ref{psi2phia}) are the inverse transformation of Eq.~(\ref{ex1}).
To summarise, we have shown that our invertibility conditions put strong constraints on the transformation, and once they are imposed it indeed becomes invertible. 
Note that there are transformations more general than (\ref{phipsi_linear}), and our invertibility conditions can be applied to examine their invertibility.

\section{Summary and discussions}

In this {\it Letter}, we formulated explicitly the necessary and
sufficient conditions for the local invertibility of a field
transformation involving derivatives.  Our result generalises the
well-known inverse function theorem.  In order to address this problem,
we used the method of characteristics for differential equations,
regarding such a transformation as differential equations 
that give new variables in terms of original ones.
For the invertibility of a
transformation, we require that the coefficients of the highest-order
derivatives in such differential equations degenerate.
With this condition
the order of differential equations can be reduced once a specific combination of equations is taken.
This procedure needs to be repeated
until all the derivatives are effectively eliminated from the system of
equations.  
It is necessary also that the characteristic matrix after the above manipulation is non-degenerate, 
resulting in the conditions that guarantee the inverse function theorem to work.
As the simplest nontrivial case, we studied in detail the
case with two fields involving first derivatives (see
Eq.~(\ref{phipsi})), and we 
found the necessary conditions for the invertibility 
(\ref{aVU}), (\ref{nBm}) and (\ref{nBU}).
We confirmed that they are also sufficient conditions for the invertibility.

The study of the invertibility conditions presented in this paper can be directly applied 
to the case of classical mechanics, in which case the fields $\phi_i(x^{\mu})$ and $\psi_a(x^{\mu})$ should be replaced 
by $q_a(t)$ and $\tilde{q}_b(t)$ and the transformation 
$q^a = q^a\left(\tilde{q}^b,d\tilde{q}^b/dt\right)$ should be considered instead of~(\ref{phipsi1}).
The details of this study will be discussed in the forthcoming paper.

Along with the general analysis, we also presented a concrete non-trivial example to show how our conditions work for invertibility of a transformation, where the standard inverse function theorem does not apply.
This example, in particular, shows how powerful the conditions formulated in this {\it Letter} are:
without the use of them one would not be able to find whether the non-trivial transformation used in this example is invertible or not.

Our findings may have many interesting applications in various fields. 
The use of invertible transformations with derivative terms greatly broadens the ``standard'' non-derivative transformations.
In our future work we will extend the presented analysis to include an arbitrary number of fields 
as well as an arbitrary order of derivatives in the transformation. 
They will include transformation of the metric {\it \`a la} disformal transformation,
which has been recently a very popular approach in gravity theories.
In particular, we will study possible extensions of disformal transformation.

\section{Acknowledgments}

E.B. acknowledges support from the research programs ``Projet 80$|$PRIME
CNRS'' and PRC CNRS/RFBR (2018--2020) n\textsuperscript{o}1985.  This
work is partially supported by the JSPS KAKENHI Grant Numbers
JP17K1428(K.I.), JP17H0109(K.I.), JP18K03623 (N.T.), JP18K18764(M.Y.),
by the MEXT Grant-in-Aid for Scientific Research on Innovative Areas
Nos.15H05888, 18H04579 (M.Y.), and by the Mitsubishi Foundation (M.Y.).
K.I.\ is supported by Japan-Korea Bilateral Joint Research Projects
(JSPS-NRF collaboration) String Axion Cosmology.  M.Y.\ is supported by
JSPS and NRF under the Japan-Korea Basic Scientific Cooperation Program
and would like to thank the participants attending the JSPS and NRF
conference for useful comments.


\begin{thebibliography}{99}

\bibitem{Nambu:1960tm} 
  Y.~Nambu,
  Phys.\ Rev.\  {\bf 117}, 648 (1960).
  doi:10.1103/PhysRev.117.648

\bibitem{Goldstone:1961eq} 
  J.~Goldstone,
  Nuovo Cim.\  {\bf 19}, 154 (1961).
  doi:10.1007/BF02812722

\bibitem{Bogoljubov:1958}
N.~N.~Bogoljubov, 
Nuovo Cim.\  {\bf 7}, 794 (1958).
doi.org/10.1007/BF02745585 

\bibitem{Valatin:1958}
J.~G.~Valatin, 
Nuovo Cim.\ {\bf 7}, 843 (1958). 
doi.org/10.1007/BF02745589 

\bibitem{York:1971hw} 
  J.~W.~York, Jr.,
  Phys.\ Rev.\ Lett.\  {\bf 26}, 1656 (1971).
  doi:10.1103/PhysRevLett.26.1656

\bibitem{York:1972sj} 
  J.~W.~York, Jr.,
  Phys.\ Rev.\ Lett.\  {\bf 28}, 1082 (1972).
  doi:10.1103/PhysRevLett.28.1082

\bibitem{Bekenstein:1992pj} 
  J.~D.~Bekenstein,
  Phys.\ Rev.\ D {\bf 48}, 3641 (1993)
  doi:10.1103/PhysRevD.48.3641
  [gr-qc/9211017].
  
\bibitem{Zumalacarregui:2013pma}
  M.~Zumalacarregui and J.~Garcia-Bellido,
  Phys.\ Rev.\ D {\bf 89} (2014) 064046
  doi:10.1103/PhysRevD.89.064046
  [arXiv:1308.4685 [gr-qc]].


\bibitem{Takahashi:2017zgr} 
  K.~Takahashi, H.~Motohashi, T.~Suyama and T.~Kobayashi,
  Phys.\ Rev.\ D {\bf 95}, no. 8, 084053 (2017)
  doi:10.1103/PhysRevD.95.084053
  [arXiv:1702.01849 [gr-qc]].

\bibitem{CH:1962}
R. Courant and D. Hilbert, 
Methods of Mathematical Physics, Vol.2, Wiley-Interscience, Inc., New York (1962).

\bibitem{Ong:2013qja} 
  Y.~C.~Ong, K.~Izumi, J.~M.~Nester and P.~Chen,
  Phys.\ Rev.\ D {\bf 88}, 024019 (2013)
  doi:10.1103/PhysRevD.88.024019
  [arXiv:1303.0993 [gr-qc]].

\bibitem{Izumi:2013poa} 
  K.~Izumi and Y.~C.~Ong,
  Class.\ Quant.\ Grav.\  {\bf 30}, 184008 (2013)
  doi:10.1088/0264-9381/30/18/184008
  [arXiv:1304.0211 [hep-th]].

\bibitem{Izumi:2014loa} 
  K.~Izumi,
  Phys.\ Rev.\ D {\bf 90}, no. 4, 044037 (2014)
  doi:10.1103/PhysRevD.90.044037
  [arXiv:1406.0677 [gr-qc]].

\bibitem{Reall:2014pwa} 
  H.~Reall, N.~Tanahashi and B.~Way,
  Class.\ Quant.\ Grav.\  {\bf 31}, 205005 (2014)
  doi:10.1088/0264-9381/31/20/205005
  [arXiv:1406.3379 [hep-th]].


\end{thebibliography}
\end{document}